\documentclass[sigconf, nonacm]{acmart}

\usepackage{amsmath}
\usepackage{verbatim}

\usepackage{xurl}
\hypersetup{breaklinks=true}

\newcommand\vldbdoi{XX.XX/XXX.XX}
\newcommand\vldbpages{XXX-XXX}
\newcommand\vldbvolume{14}
\newcommand\vldbissue{1}
\newcommand\vldbyear{2020}
\newcommand\vldbauthors{\authors}
\newcommand\vldbtitle{\shorttitle} 
\newcommand\vldbavailabilityurl{URL_TO_YOUR_ARTIFACTS}
\newcommand\vldbpagestyle{plain} 

\begin{document}





\title{Solving Data Quality Problems with Desbordante: a Demo}


\author{
George Chernishev$^{1,2}$, 
Michael Polyntsov$^2$, 
Anton Chizhov$^2$, 
Kirill Stupakov$^{1,2}$,
Ilya Shchuckin$^2$,
Alexander Smirnov$^2$, 
Maxim Strutovsky$^{1,2}$, 
Alexey Shlyonskikh$^2$, 
Mikhail Firsov$^2$, 
Stepan Manannikov$^2$,
Nikita Bobrov$^2$,
Daniil Goncharov$^2$,
Ilia Barutkin$^2$,
Vladislav Shalnev$^2$,
Kirill Muraviev$^2$,
Anna Rakhmukova$^2$,
Dmitriy Shcheka$^2$,
Anton Chernikov$^2$,
Mikhail Vyrodov$^2$,
Yaroslav Kurbatov$^2$,
Maxim Fofanov$^2$,
Sergei Belokonnyi$^2$,
Pavel Anosov$^2$,
Arthur Saliou$^2$,
Eduard Gaisin$^2$,
Kirill Smirnov$^2$}

\affiliation{%
  \institution{$^1$Universe Data LLC, $^2$Saint-Petersburg University}
  \city{Saint-Petersburg}
  \state{Russia}
}

\email{
chernishev,
polyntsov.m,
anton.i.chizhov,
kirill.v.stupakov,
shchuckinilya, 
alexander.a.smirnovv,
strutovsky.m.a,
}

\email{
shlyonskikh.alexey,
mikhail.a.firsov,
manannikov.st,
nikita.v.bobrov,
goncharov.y.daniil,
ilia.d.barutkin,
vladislav.a.shalnevv,
}

\email{
kirill.i.muravev,
rakhmukova.anna,
dmitriy.v.shcheka,
chernikov.a.anton,
mikhail.v.vyrodov,
yaroslav.a.kurbatov,
max.fofanov,
}

\email{
belokoniy,
pavel.i.anosov,
arthur.salio,
edu.gaisin,
kirill.k.smirnov
@gmail.com
}

\begin{abstract}

Data profiling is an essential process in modern data-driven industries. One of its critical components is the discovery and validation of complex statistics, including functional dependencies, data constraints, association rules, and others.

However, most existing data profiling systems that focus on complex statistics do not provide proper integration with the tools used by contemporary data scientists. This creates a significant barrier to the adoption of these tools in the industry. Moreover, existing systems were not created with industrial-grade workloads in mind. Finally, they do not aim to provide descriptive explanations, i.e. why a given pattern is not found. It is a significant issue as it is essential to understand the underlying reasons for a specific pattern's absence to make informed decisions based on the data.

Because of that, these patterns are effectively rest in thin air: their application scope is rather limited, they are rarely used by the broader public. At the same time, as we are going to demonstrate in this presentation, complex statistics can be efficiently used to solve many classic data quality problems.

Desbordante is an open-source data profiler that aims to close this gap. It is built with emphasis on industrial application: it is efficient, scalable, resilient to crashes, and provides explanations. Furthermore, it provides seamless Python integration by offloading various costly operations to the C++ core, not only mining.

In this demonstration, we show several scenarios that allow end users to solve different data quality problems. Namely, we showcase typo detection, data deduplication, and data anomaly detection scenarios.

\end{abstract}

\maketitle

\pagestyle{\vldbpagestyle}
\begingroup\small\noindent\raggedright\textbf{PVLDB Reference Format:}\\
\vldbauthors. \vldbtitle. PVLDB, \vldbvolume(\vldbissue): \vldbpages, \vldbyear.\\
\href{https://doi.org/\vldbdoi}{doi:\vldbdoi}
\endgroup
\begingroup
\renewcommand\thefootnote{}\footnote{\noindent
This work is licensed under the Creative Commons BY-NC-ND 4.0 International License. Visit \url{https://creativecommons.org/licenses/by-nc-nd/4.0/} to view a copy of this license. For any use beyond those covered by this license, obtain permission by emailing \href{mailto:info@vldb.org}{info@vldb.org}. Copyright is held by the owner/author(s). Publication rights licensed to the VLDB Endowment. \\
\raggedright Proceedings of the VLDB Endowment, Vol. \vldbvolume, No. \vldbissue\ %
ISSN 2150-8097. \\
\href{https://doi.org/\vldbdoi}{doi:\vldbdoi} \\
}\addtocounter{footnote}{-1}\endgroup

\ifdefempty{\vldbavailabilityurl}{}{
\vspace{.3cm}
\begingroup\small\noindent\raggedright\textbf{PVLDB Artifact Availability:}\\
The source code, data, and/or other artifacts have been made available at \url{https://github.com/Mstrutov/Desbordante}.
\endgroup
}

\section{Introduction}

Data profiling subsumes a broad range of techniques of processing data sets and extracting metadata. Metadata comes in many forms, and it is crucial for interpretation of data and drawing unambiguous conclusions about its nature. Dozens of tools have emerged during the decades of data profiling evolution, and each of them approaches profiling tasks with both well-known methods or an original research solutions to a specific problem. 

Since there are so many tools, there are many ways to classify them: by their availability (commercial, free-to-use, open-source), interactivity, types of data that can be processed (tabular, graph, heterogeneous, etc.), and many more. Since there is no canonical attribute to construct a data profiling tool classification upon, we need to define one for this paper. From now on, we consider tools from the angle of the \textit{complexity of extracted metadata}. This characteristic well-defines the profiling workflow itself, along with the set of tasks that can be solved. We believe that the extent of metadata complexity expresses a dichotomy of data profiling approaches as \textit{naive} and \textit{science-intensive}. 

Naive data profiling manipulates easy-to-extract metadata such as number of rows/columns/missing cells in a table, maximum/mini\-mum/average value in a column, various statistical calculations. Most commercial and free-to-use systems offer this functionality. 

Science-intensive profiling, in turn, captures hidden knowledge that can be extracted only by advanced data mining techniques. Such knowledge is represented by structures which we will refer to as \textit{primitives}~--- it can be any type of database dependency, association rule, algebraic constraint, inferred semantic data type, etc. which can be included into a complex data profiling workflow. The actual value for the profiling process is provided by an \textit{instance} of a primitive, which is simply an object of some primitive's type, e.g. a specific functional dependency. 

The presence of primitives within a toolkit of a contemporary data profiler engine is a phenomenon rather than a rule. Moreover, only a handful of them allow the user to actually mine primitives and include them into the workflow~\cite{10.14778/2824032.2824086, 10.14778/3476311.3476339}. However, even the ones which have primitive discovery do not guarantee the best possible performance due to specifics of system implementations. It leaves room for improvement in terms of run time and memory consumption, which can be accomplished by applying engineering and architectural solutions that have not been considered before.

In this paper, we demonstrate Desbordante (Spanish for \textit{boundless})~--- a \textit{science-intensive}, \textit{high-performance} and \textit{open-source} data profiling tool implemented in C++. To the best of our knowledge, Desbordante is currently the only profiler which possesses these three characteristics. 

Desbordante comes with web, console, and Python interfaces. The latter makes calling primitive discovery and validation tasks from Python programs possible. Desbordante provides users with the opportunity to seamlessly integrate it with other data science libraries, including Pandas, creating a wealth of possibilities for data scientists. Our primary objective is to empower end-users to apply primitive discovery and validation techniques in a range of ad-hoc scenarios, with the ultimate aim of solving real-life problems. A more detailed description of Desbordante can be found in paper~\cite{DBLP:journals/corr/abs-2301-05965}.

In this demonstration, we present three scenarios solving common data profiling tasks, and we also showcase the Web UI of Desbordante. 
The first one concerns data cleaning (typo detection), the second implements deduplication, and the last one demonstrates a scenario for anomaly detection. 

The ad-hoc solutions are implemented in Python, contain modest amount of code, and offer high result quality. For all of them, Desbordante provides the core functionality. At the same time, these scenarios should not be considered perfect solutions, but instead, they are more of a playground, illustrating the principal possibility of incorporating primitives into data science workflows and offering rapid prototyping opportunities.

During the demonstration a user will be presented Python notebooks (links are in the corresponding sections) and Streamlit application (link\footnote{\url{https://desbordante.unidata-platform.ru/streamlit}}) which implement each procedure step by step. Then, an example dataset will be loaded and its playthrough will be shown. Also, the impact of various parameters on the script will be demonstrated.

\section{Desbordante}

\subsection{Tool Overview}

\textbf{General.} Desbordante supports two different types of tasks, and, therefore, algorithms: 1) Discovery: find all holding instances of specified primitive over specified dataset; 2) Validation: given a primitive instance, determine whether it holds over the specified dataset, provide additional information about what prevents it from holding otherwise. 

Desbordante covers a fairly broad range of primitives and algorithms such as functional/inclusion/graph dependencies, association rules, algebraic constraints, etc. The full list can be seen in paper~\cite{DBLP:journals/corr/abs-2301-05965}. Desbordante comes with a web, console, and Python interfaces. 

Currently, there are only two open-source systems that specialize in science-intensive profiling functionality: Me\-ta\-no\-me~\cite{10.14778/2824032.2824086} and OpenClean~\cite{10.14778/3476311.3476339}. Desbordante is different and offers more:

\textbf{1.} Both of them share the same Java-based core and therefore, both of them are performance-constrained. On the other hand, Desbordante is built with intent to employ low-level optimizations such as vectorization, cache-conscious programming and so on.

\textbf{2.} The web application of Desbordante is built with emphasis on industrial application in a multi-user environment. It is efficient, resilient to crashes, and scalable. Resilience is achieved by extensive use of containerization, and scalability is based on replication of containers.

\textbf{3.} It offers result explainability: Desbordante not only reports whether a given instance of primitive holds or not, but also points out what prevents it from holding via special screens. 

\textbf{4.} Similarly to Desbordante, OpenClean also provides Python integration. However, it has one serious drawback~--- the shallow integration with Metanome. OpenClean's primitive mining functionality is provided by a standalone package which initiates a subprocess for running Metanome JAR files. Therefore, OpenClean can acquire only that information that was provided by Metanome's algorithms. It is very restrictive as it requires to reproduce costly procedures inside Python code in order to obtain additional information, which is frequently needed for providing explanations or organizing complex workflows. On the other hand, Desbordante implements many additional methods inside C++ core.

\textbf{Implementation.} The core of Desbordante is a C++ library containing all auxiliary data structures needed for primitive discovery algorithms, the algorithms themselves and all required surrounding infrastructure. The library provides an API for running the algorithms and obtaining their results which is used by the back-end of the web application. There is also an additional library version with Python bindings, so that all Desbordante features can be used from Python programs.

Unlike all existing open-source solutions, the discovery part of Desbordante is fully implemented in modern C++. This makes Desbordante significantly faster and makes it consume less memory.

For example, for functional dependency discovery task the speed-up ranged~\cite{9435469} from 1.19 to 3.43 times and was 2.12 on average, compared to Metanome. While the numbers are not really high, it is still a significant result for such a computationally expensive problem.

It is important to note that for this primitive, we have not tapped into the tuning potential of the C++ implementation. Sophisticated techniques (e.g., vectorization) were not used, no source code profiling was done, standard data structures and libraries were used, etc. Currently, Desbordante relies on default C++ and Boost data structures, and we have not tuned their parameters. Desbordante does not rely on custom memory management libraries (allocators), but instead uses the C++ default. It is a well-known fact~\cite{MemoryAllocators} that using a special allocator is a simple yet efficient way to improve performance of C++ programs. 

Recently, we have performed a pen test of advanced techniques for implementing inclusion dependency discovery algorithms~--- Faida and Spider~\cite{fruct23}. In this case, we used a simple vectorization approach, custom hash tables, and we have profiled code. As the result, we have managed to obtain 3x--7x run time improvement over Metanome. Therefore, it is possible to achieve major performance improvements by designing efficient, cache-conscious C++ implementations of algorithms.

Another significant benefit of C++ implementation is the reduction of memory consumption. For example, the memory footprint of Pyro, a functional dependency mining algorithm~\cite{10.14778/3192965.3192968}, is approximately two times lower in Desbordante than in Metanome~\cite{9435469}. This is crucial since many primitive discovery algorithms are memory-bound~\cite{10.1145/2882903.2915203}. Thus, reducing memory footprint enables the processing of larger datasets.

\subsection{Python Interface}

Desbordante features can be accessed from within Python programs by employing the Desbordante Python library. The library is implemented in the form of Python bindings to the interface of Desbordante C++ core library, using pybind11~\cite{pybind11}. The workflow with the bindings is as follows: first, the user selects an algorithm which can be achieved by creating an algorithm object in Python; then configures it, calls the \texttt{Execute} method and retrieves results through appropriate methods specific to the type of an algorithm. Despite the fact that the procedure is the same as when working with a web UI, Python bindings offer much more flexibility. Relational data processing algorithms accept Pandas dataframes as input, allowing the user to conveniently preprocess data before mining primitives. The results of an algorithm can be converted to basic Python types such as lists and tuples, so it is easy to perform additional filtering and other further processing techniques with the powerful facilities Python provides. We will leverage such techniques in the following demonstration scenarios. 

\section{Demonstration Scenarios}

\subsection{Scenario 1: Typo Detection}

The first scenario addresses a frequent use-case of a table which needs to be cleaned. For this purpose, we use algorithms for discovery of exact and approximate functional dependencies. In the first approximation, the approach is as follows: 1) Desbordante discovers all minimal dependencies which ``almost'' hold, 2) the user selects a particular dependency for inspection, 3) Desbordante presents ``clusters'' in which the exact FD is violated. For a fixed functional dependency LHS $\rightarrow$ RHS, a cluster is a row subset  which has identical values in LHS, but different in the RHS. It is possible to locate typos by looking into these clusters, given the fact that they are rarely occurring.

The tricky part is the selection of the threshold for ``almost''-holding dependencies. The notion of AFD~\cite{10.14778/3192965.3192968} provides no straightforward means to associate the threshold with the number of tuples. Therefore, a user has to experiment with the \texttt{threshold} parameter. Another issue is the potentially big amount of clusters if an exact FD does not hold semantically on the schema. To address it, we have two additional parameters~--- \texttt{radius} and \texttt{ratio}. The idea is the following: a cluster is shown if the share of records residing inside the \texttt{radius} is less than \texttt{ratio}. The \texttt{radius} shows how close a row to ``central'' record should be to consider it containing a typo. Distance is calculated for RHS values and can be computed using the Levenshtein metric for the string type or using the $L_2$ for numeric types. The resulting script can be found by the following link\footnote{\url{https://colab.research.google.com/drive/1h5mQAIIxSb6Sgc_Ep8AYZlgt4BGXN6A9}}. Note that this scenario was not only implemented as a Python script, but it also has a Web UI visualization, which will be briefly demonstrated in the last Scenario.

\subsection{Scenario 2: Data Deduplication}

The second scenario concerns data deduplication for tabular datasets. Similarly to the previous scenario, it is a multi-step procedure, but each step requires user input. The approach is inspired by the Example 6 of the SSJoin paper~\cite{1617373} and our implementation relies on the approximate functional dependency discovery primitive.

The workflow is as follows: 1) Desbordante discovers all approximate functional dependencies $X_i \longrightarrow X_j$, where $X_i, X_j$ are single attributes, 2) for each LHS $X_i$ the Python script groups all the respective RHSs $X_j$, so we obtain $n$ following records:
\begin{align*} 
X_1 &\longrightarrow \{ X^1_{1}, X^1_{2}, \ldots, X^1_{i_1} \} \\ 
X_2 &\longrightarrow \{ X^2_{1}, X^2_{2}, \ldots, X^2_{i_2} \} \\
&\ldots \\
X_n &\longrightarrow \{ X^n_{1}, X^n_{2}, \ldots, X^n_{i_n} \},
\end{align*}
3) the user picks $X_k$ by selecting the largest list of $X^k_{1 \ldots i_k}$, and at the same time leaving out $X_k$ which are artificial primary keys of the considered table, 4) the Python script sorts the source table by the values in this list of attributes, and then 5) a sorted-neighborhood~\cite{DBLP:journals/datamine/HernandezS98}-like procedure is run. On this step, a sliding window goes over table tuples and duplicates are detected. We consider two records as duplicates if they match~\cite{1617373} on at least \texttt{k} attributes, which is set by the user at start. For each detected duplicate pair the user can select the one to keep and which attributes to copy from the other. 

This approach is based on the following considerations: 1) we try to determine a non-surrogate attribute which is as close to a key as possible (LHS of FD selected on step 3), and 2) the resulting RHS list $X^k_{1 \ldots i_k}$ represents dependent attributes which contain data quality issues (typos, missing values). Then, sorting and iteration over data is used to eliminate key-dependent attribute inconsistencies. Despite being based on rather simple heuristics, this procedure yields surprisingly good results and has short and comprehensible source code. This script can be found at the following link\footnote{\url{https://colab.research.google.com/drive/1zX2OLX3K-XKw-Nz7e9YOUSGFt_shiNAC}}.

\subsection{Scenario 3: Anomaly Detection}

The following scenario was devised in order to demonstrate the focus of Desbordante on explainability and data exploration. The workflow represents the mine-explore-validate cycle in which the user revises knowledge obtained earlier. For this, we model a dynamic environment which does not guarantee that some knowledge will persist as the new data comes by.

The proposed user activity includes the following steps: 1) to get to know the available fraction of data $D_1$ of schema $R$, the user executes Desbordante's process for mining exact functional dependencies and obtains a set of FDs $FD(D_1)$, which is also a canonical set $FD_\mathcal{C}$, 2) next, as the new partition of data $D_2$ becomes available, the user runs the same mining process to compare $FD(D_1)$ and $FD(D_2)$ and then repeats the process, 3) for some partition $D_i$ user notices that $FD_\mathcal{C} \setminus FD(D_i) = \{X \rightarrow Y\}$, 4) the user runs on $D_i$ a series of tasks that mine approximate FDs with different error thresholds, but fails to detect AFD $X \rightarrow Y$, 5) the user chooses some reasonable ``distance'' $d$ for relaxation of FD, runs the metric functional dependency (MFD)~\cite{MFD} validation processes for parameter $p$ in range $[1; d]$, and picks the smallest $p_j$ such that MFD $X \xrightarrow{\Psi_{err(0)}} Y_{\max_{s \in \pi_{X} \Delta_{\phi}(s[Y]) \leq p_j}}$ holds, 6) the user assigns $FD(D_i)$ to $FD_\mathcal{C}$ and adds found MFD to $MFD_\mathcal{C}$.

As all mining and validation processes are fully automated by Desbordante, this scenario still requires a domain expert who can competently define the range of possible values for the distance parameter. The range depends on the data type of the RHS attribute, its observable values and basic statistics. For example, in the script with the scenario\footnote{\url{https://colab.research.google.com/drive/1hgF8idXi1-U4ZOR0fAmdbfbhltgEJecR}}, the user decides to set one standard deviation as the upper bound of the range, and they also set 1 as the step value because the attribute type is integer. Obviously, an MFD validated in such a way is not always minimal, since the distance parameter value may not belong to the set of possible values, but rather lies somewhere in-between. Nevertheless, the demonstrated process can help to understand whether presented changes in $FD_\mathcal{C}$ are a result of data anomaly or they are a new norm which must be taken into account.

\subsection{Web UI}

The final demonstration does not cover a particular use-case, but instead showcases the Web version of Desbordante. 

\textbf{Primitive instance browsing.} First of all, the rich Web UI offers users a convenient way to browse discovered instances of primitives. A user can sort them using various parameters, filter them with regular expressions, and so on.

\textbf{Tuning knobs.} Next, our users are provided with various tuning knobs that will govern primitive discovery process. 
A user may wish to set up the maximum left-hand side length, the error threshold for approximate primitives, the number of threads, and so on. The overall list is very large, and it can be cumbersome to define it using the command line interface. The Web UI of Desbordante offers screens with a comfortable way for setting up various parameters for tasks.

\textbf{Explanation screens.} Desbordante offers primitive validation, which not only reports whether a given instance of primitive holds or not, but also points out what prevents it from holding via the use of special screens. It is essential to provide such information since this is important knowledge about the explored data. It can indicate errors in data and point out problematic records. Therefore, there is a need for screens that will provide this information. Desbordante aims to provide such screen for each validation primitive.

\textbf{Web versions of scenarios.} Some Python scenarios can be implemented in the web version of Desbordante. The Web UI allows to provide rich user interactions, which is almost impossible to achieve in the text-only Python programs. On the other hand, the CLI version does not yet provide such scenarios at all, since they require extensive interactivity. Therefore, currently the Web version and the Python interface are the only ways to implement interactive scenarios. Most modern applications use the Web interface as their primary way of user interaction. Since Desbordante is in the prototype phase, the Web interface is not our primary focus. Instead, we concentrate on the science-intensive parts of data profiling. Turning to scenarios, our idea is to allow users to experiment and build their own scenarios in Python. Most popular and proven ones will be added to the web version and provided with a user-friendly interface. However, they require significant effort to implement in the Web UI. Since our resources are limited, for now we put online only a single scenario~--- the web version of the typo detection script, which can be found at the following link\footnote{\url{https://desbordante.unidata-platform.ru/}}.

\section*{Acknowledgments}

We would like to thank Nikita Talalay and Bulat Biktagirov for their contribution to the project. We would also like to thank Anna Smirnova for her help with the preparation of this paper.

\bibliographystyle{ACM-Reference-Format}
\bibliography{our-short}

\end{document}